\documentclass[reprint, 
superscriptaddress, 
nofootinbib,
aps, 
prd,
preprintnumbers %added to show preprint numbers 
]{revtex4-2}%Delete onecolumn and 11pt later.

\pdfoutput=1
\usepackage{mathtools, amsfonts, amsthm, latexsym, amssymb, dsfont, mathrsfs}
\usepackage[T1]{fontenc}
\usepackage{newtxtext}
\usepackage{physics2}
\usephysicsmodule{ab}
\usepackage{braket}
\usepackage{color}
\usepackage{bm}

\usepackage{tikz}
\usetikzlibrary{decorations.markings}

\usepackage{hyperref}
\hypersetup{colorlinks=true, citecolor=blue, urlcolor=blue}
\usepackage{cleveref}

\usepackage{titlesec}
\titlespacing{\section}
    {0pt}% left
    {5pt}% before-sep
    {3pt}% after-sep
\titlespacing{\subsection}
    {0pt}% left
    {5pt}% before-sep
    {3pt}% after-sep

\AtBeginDocument{
  \abovedisplayskip     =0.7\abovedisplayskip
  \abovedisplayshortskip=0.7\abovedisplayshortskip
  \belowdisplayskip     =0.7\belowdisplayskip
  \belowdisplayshortskip=0.7\belowdisplayshortskip}

\makeatletter
\renewcommand\onecolumngrid{%Changed so that footnote is also in onecolumngrid in appendix
  \do@columngrid{one}{\@ne}%
  \def\set@footnotewidth{\onecolumngrid}%
  \def\footnoterule{\kern-6pt\hrule width 1.5in\kern6pt}%
}
\makeatother

\newcommand{\del}{\partial}
\newcommand{\nn}{\nonumber\\}

\newcommand{\order}[1]{\mathcal{O} \ab( #1 )}
\newcommand{\rS}{\mathrm{S}}
\newcommand{\rH}{\mathrm{H}}

\begin{document}
\title{Asymptotic symmetry and confinement in three-dimensional QED}
\author{Keito Shimizu}
\email{kate@gauge.scphys.kyoto-u.ac.jp}
\affiliation{Department of Physics, Kyoto University, Kitashirakawa-Oiwakecho, Kyoto 606-8502, Japan}
\author{Sotaro Sugishita}
\email{sotaro_at_gauge.scphys.kyoto-u.ac.jp}
\affiliation{Department of Physics, Kyoto University, Kitashirakawa-Oiwakecho, Kyoto 606-8502, Japan}
%\date{\today}
\preprint{KUNS-3041}
% \begin{flushright}
% KUNS-3041
% \end{flushright}

\begin{abstract}
We investigate the asymptotic symmetries of quantum electrodynamics (QED) in three dimensions, demonstrating that their actions on asymptotic states are trivial under the assumption of confinement. 
\end{abstract}
\maketitle
%, unlike four-dimensional QED.
%We examine their actions on asymptotic states under the assumption of confinement and find them trivial.
%We conjecture that the triviality of asymptotic symmetries holds in confined gauge theories.

\section{Introduction}
Over the past decade, the infrared (IR) structure of gauge and gravitational theories on asymptotically flat spacetime has been studied through \textit{asymptotic symmetries} \cite{Strominger:2013lka, Strominger:2013jfa, Cachazo:2014fwa, He:2014cra, He:2014laa, Campiglia:2014yka, Kapec:2014opa, Kapec:2014zla, Larkoski:2014bxa, Liu:2014vva, Lysov:2014csa, Strominger:2014pwa, Kapec:2015vwa, He:2015zea, Campiglia:2015qka, Kapec:2015ena, Campiglia:2015kxa, Dumitrescu:2015fej, Strominger:2015bla, Pasterski:2015tva, Campiglia:2016hvg, Strominger:2017zoo, Campiglia:2017dpg, Pate:2017vwa, Hamada:2017atr, Pate:2017fgt, Hamada:2018cjj, Hirai:2018ijc, Campiglia:2018see, Francia:2018jtb, 
Henneaux:2018mgn}.
For example, soft theorems concerning low-energy scatterings of massless particles were found to be equivalent to the Ward-Takahashi identity of asymptotic symmetries.
Moreover, memory effects were revealed as the conservation law of asymptotic charges.
Asymptotic symmetries are also relevant to numerous studies including IR divergences \cite{Mirbabayi:2016axw, Kapec:2017tkm, Carney:2018ygh, Neuenfeld:2018fdw, Hirai:2019gio, Gonzo:2019fai, Furugori:2020vdl, Hirai:2020kzx, Hirai:2022yqw}, amplitudes \cite{Chen:2014xoa, Rao:2016tgx, Cachazo:2016njl, Rodina:2018pcb, Moult:2019mog, Derda:2024jvo}, gravitational waves \cite{Zeldovich:1974gvh, Braginsky:1985vlg, Braginsky:1987kwo,  Lasky:2016knh, Zhang:2017rno, Zhang:2017geq, DeLuca:2024bpt} and the celestial holography \cite{Pasterski:2016qvg, Pasterski:2017ylz, Cardona:2017keg, Pasterski:2017kqt, Arkani-Hamed:2020gyp, Donnay:2020guq, Raclariu:2021zjz, Pasterski:2021rjz, Sleight:2023ojm, Hao:2023wln, Banerjee:2024hvb, Melton:2024akx}.

Despite many works on low-energy aspects of gauge theories in terms of asymptotic symmetries, previous studies have paid little attention to confinement, a crucial aspect of the low-energy behavior of gauge theories.\footnote{
As far as we know, \cite{Tanzi:2020fmt} is one of the few exceptions.}
In gauge theories, the asymptotic symmetries, which are a class of large gauge transformations, can be physical symmetry (not gauge redundancy), unlike small gauge transformations.
The action of asymptotic symmetries depends on the asymptotic structure (i.e. asymptotic physical phase space) of the theory under consideration. 
Since confinement prevents charged particles from existing independently, it is reasonable to anticipate that properties of asymptotic symmetries in confinement theories such as four-dimensional quantum chromodynamics (QCD) are drastically different from those in deconfined theories such as four-dimensional quantum electrodynamics (QED).
Indeed, it has been discussed and expected that a kind of large gauge symmetry\footnote{The large gauge symmetry discussed in \cite{Ferrari:1971at, Hata:1981nd} is slightly different from the asymptotic symmetry in recent papers because the charges for the former transformation are not well-defined.} is spontaneously broken in QED$_4$ \cite{Ferrari:1971at} and it is not broken in QCD$_4$ \cite{Hata:1981nd}.

In QED$_4$, the action of the asymptotic symmetries on asymptotic states has already been investigated in various literature, e.g., \cite{He:2014cra, Campiglia:2015qka, Kapec:2015ena, Hirai:2018ijc}.
The asymptotic charges defined on future/past infinities are the Noether charges associated with large U(1) gauge transformations 
with a class of gauge parameters $\alpha(x)$ and divided into two parts, the hard and soft parts:
\begin{align}
\label{eq:Q-QED4}
    Q^{\pm}[\alpha] = Q_{\mathrm{hard}}^{\pm}[\alpha] + Q_{\mathrm{soft}}^\pm[\alpha].
\end{align}
The hard part is determined from the asymptotic data of charged particles. 
The soft part is related to long-range behaviors of gauge fields and essentially acts only on soft photons.
The sum \eqref{eq:Q-QED4} is conserved as $Q^+=Q^-$, not each part.
Thus, the mixing of the hard and soft parts is allowed, and it leads to the soft photon theorem \cite{Weinberg:1965nx} and the electromagnetic memory effect \cite{Bieri:2013hqa, Susskind:2015hpa}.
In addition, the symmetry is spontaneously broken, and consequently the spectrum of QED$_4$ is infinitely degenerate.
More precisely, the Hilbert space is decomposed into the superselection sectors determined by the value of $Q^\pm$, and states in each sector have dresses of soft photons (see also \cite{Chung:1965zza, Kibble:1968sfb, Kibble:1969ip, Kibble:1969ep, Kibble:1969kd}). Such dressed states are inevitable to realize the conservation law $Q^+=Q^-$ and to obtain IR-safe $S$-matrix elements \cite{Mirbabayi:2016axw, Kapec:2017tkm, Hirai:2020kzx, Hirai:2022yqw}.

Does a similar structure hold in confined gauge theories? 
We conjecture that the answer is no; \textit{the asymptotic symmetries act on asymptotic states trivially}.
This is because in confined theories, there is no charged particle in asymptotic states and thus there is no way for them to dress up with soft massless gauge particles.
Consequently, degeneracies arising from dressing are absent, resulting in the trivial action of asymptotic symmetries.
In other words, the asymptotic symmetries are a gauge redundancy in confined theories.

It sounds trivial for gapped theories -- such as QCD$_4$, which is believed to be gapped -- because the absence of massless particles in the spectrum implies that there is no freedom contributing to the soft part $Q_{\mathrm{soft}}^\pm$.
We speculate that our conjecture holds even in gapless theories. 

As a non-trivial support for this conjecture, we consider (2+1)-dimensional QED in this paper (we refer \cite{Polyakov:1976fu, Gopfert:1981er, Burden:1991uh, Roberts:1994dr, Grignani:1995iv, Maris:1995ns} for references of QED$_3$). 
In this theory, charged particles are confined, while massless photons can exist. 
We investigate the connection between the confinement and the asymptotic symmetries in this theory.
Specifically, we construct the asymptotic charges and demonstrate that their actions on asymptotic states are trivial.
We note here that asymptotic symmetries in three-dimensional (asymptotically) flat spacetime are also discussed, e.g., in \cite{Ashtekar:1996cd, Barnich:2006av, Barnich:2010eb, Cotler:2024cia}.

\section{Electromagnetism in three dimensions}
We consider the Maxwell action in 2+1-dimensions minimally coupled with charged fields as
\begin{align}
    \mathcal{L} = -\frac{1}{4}F_{\mu\nu}F^{\mu\nu} + \mathcal{L}_{\mathrm{charge}},
\end{align}
where $\mathcal{L}_{\mathrm{charge}}$ is the Lagrangian of charged particles.
In this paper, we consider only massive charged particles to avoid technical difficulties due to massless particles. 
Although we do not specify the kind of matter fields, we assume that they do not produce the mass term of photons via loop effects.\footnote{A charged Dirac fermion with mass $m$ and charge $e$ produces the Chern-Simons term at the 1-loop, and it induces the photon mass term proportional to $e^2\mathrm{sgn}(m)$. For example, the photon mass term is not induced, if we have a pair of Dirac fermions with the same charge but opposite-sign masses.}
Thus, we suppose that photons are massless at the quantum level.

We first review the classical property. 
The confinement property of this theory indeed can be seen even at the classical level because the potential between charged particles diverges as a logarithm of the distance.\footnote{We call it \textit{logarithmic confinement}. Including quantum effects, this behavior of potential may change. For example, it is expected that we have a linear potential if the gauge group is compact U(1) \cite{Polyakov:1976fu}.} 
%It is in contrast with four-dimensional QCD, in which quantum effects modify the classical $1/r$ potential into the linear potential.

The equation of motion for the gauge field is $\Box A^\mu =- j^\mu$ in the Lorenz gauge $\nabla_\mu A^\mu=0$, where $j^\mu$ is the Noether current associated with global U(1) symmetry of $\mathcal{L}_{\mathrm{charge}}$ and our convention of the metric signature is mostly plus one.
By means of Green's function which satisfies $\Box G(x-x') = -\delta^{(3)}(x-x')$, the solution is given by
\begin{align}
    A^\mu(x) = \int d^3x'\ G(x-x')j^\mu(x').
\end{align}
We here take the retarded Green function as $G(x-x')$ and the explicit formula in 2+1 dimensions is given by
\begin{align}
   & G(x-x') = \frac{\Theta(t-t'-|\bm{x}-\bm{x}'|)}{2\pi\sqrt{(t-t')^2-|\bm{x}-\bm{x}'|^2}}e^{-\varepsilon (t-t')}
\end{align}
where $\bm{x}, \bm{x}'$ represent the spatial coordinate, $\Theta(x)$ is the step function, and $\varepsilon$ is an adiabatic factor introduced to shift the poles in the energy plane, or equivalently, to realize the boundary condition corresponding to the retarded Green function.
For example, the gauge field produced by a uniformly moving charged particle (mass $m$ and charge $e$) along the trajectory $\bm{x}(t)= (\bm{p}/E) t$ is 
\begin{align}
\label{eq:gauge field by stationary particle}
    &A^\mu(x) 
    %= - \frac{ev^\mu}{2\pi\,\sqrt{1-\bm{v}^2}} \log\ab(\frac{\sqrt{(\Tilde{\bm{x}}\cdot\bm{v})^2 + \Tilde{\bm{x}}^2(1-\bm{v}^2)}}{1-\bm{v}^2}\varepsilon),
     = - \frac{e p^\mu}{2\pi m} \log\ab(\frac{\varepsilon E\sqrt{(p \cdot x)^2 + m^2 (x \cdot x)}}{m^2}),
\end{align}
where we have omitted the terms that vanish in the limit $\varepsilon \to 0$.

This result leads to the logarithmic confinement of charged particles.
The potential energy between two charged particles with distance $r$ diverges as $\log r$ in the infrared limit $r \to \infty$.
% Let us consider two stationary charged particles $Q_1$ and $Q_2$ fixed at $\bm{x}$ and $\bm{y}$ respectively.
% The potential energy between them can be read off from \eqref{eq:gauge field by stationary particle} as
% \begin{align}
%     V(r) = -\frac{Q_1Q_2}{2\pi}\log\frac{r}{r_0}\qquad (r=|\bm{x}-\bm{y}|), \label{eq:potential in two dimensional space}
% \end{align}
% where $r_0$ is an arbitrary constant which has the dimension of length.
% Obviously, this potential suffers from two kinds of divergence: ultraviolet divergence ($r\to 0$) and infrared divergence ($r\to\infty$).
% In particular, the infrared divergence affects the construction of asymptotic states.
For comparison, let us recall QED$_4$ where the potential energy is suppressed in the infrared as $1/r$.
This suppression of the potential allows us to separate two charges at a long distance and electrons can exist as independent particles in the asymptotic region.
% This suppression of the potential allows us to separate two charges at a long distance and to take any tensor products of single-particle states as asymptotic states.
% Hence, as we know, a single electron can move as an independent particle and it can be observed.
In three dimensions, however, asymptotic charged particles are prohibited to exist by the finite energy condition.
This is the logarithmic confinement of QED$_3$.

To investigate asymptotic symmetries later, let us evaluate the asymptotic behavior of the Coulomb potential \eqref{eq:gauge field by stationary particle} near the future/past null infinities $\mathscr{I}^\pm$.
For this purpose, we use the retarded/advanced Bondi coordinates by introducing $u:=t-r, v:=t+r$.
The three-dimensional Minkowski coordinates $(t,x,y)$
are related to these coordinates $(u, r, \theta)$ or $(v, r, \theta)$ as
\begin{alignat}{3}
    t&=u+r,\quad &x&=r \cos \theta, \quad &y&= r\sin \theta,\\
    t&=v-r,\quad &x&=r \cos \theta, \quad &y&= r\sin \theta,
\end{alignat}
and the Minkowski line element is mapped to 
\begin{align}
    ds^2 &= -du^2-2dudr+r^2d\theta^2,\\
    ds^2 &= -dv^2+2dvdr+r^2d\theta^2.    
\end{align}
We reach the future/past null infinity $\mathscr{I}^\pm$ by taking a large $r$ limit with $(u,\theta)$ or $(v,\theta)$ fixed, respectively. 
The behavior of the gauge field \eqref{eq:gauge field by stationary particle} near $\mathscr{I}^+$ is evaluated as
\begin{align}
    A^\mu(x) \sim - \frac{e p^\mu}{2\pi m}\log\ab(\frac{\varepsilon E (E-\bm{p}\cdot \hat{x})r}{m^2}),\label{eq:gauge field near I^+}
\end{align}
at the leading order of $r$ where $\hat{x}:=\bm{x}/r$.
Each component behaves as 
\begin{align}
\left\{
\begin{alignedat}{3}
    A_u &= A_t \sim \order{\log{r}}, \\
    A_r &= A_t + \cos{\theta}A_x + \sin{\theta}A_y \sim \order{\log{r}}, \\
    A_{\theta} &= -r\sin{\theta}A_x + r\cos{\theta}A_y \sim \order{r\log{r}}.
    \end{alignedat}
    \right.
    \label{eq:comp_gauge field near I^+}
\end{align}
Similarly near $\mathscr{I}^-$, we have
\begin{align}
    A^\mu(x) \sim  - \frac{e p^\mu}{2\pi m}\log\ab(\frac{\varepsilon E (E+\bm{p}\cdot \hat{x})r}{m^2}).\label{eq:gauge field near I^-}
\end{align}
The leading part of the gauge field \eqref{eq:gauge field near I^+} or \eqref{eq:gauge field near I^-} is independent of $u$ or $v$ respectively, and does not contribute to the soft part of asymptotic charges.

\section{Asymptotic boundary conditions and asymptotic symmetry}
What contributes to the soft part is a long-range part of the radiation modes. 
We define the radiation modes as ones with the dispersion relation of free photons\footnote{The Coulomb potential \eqref{eq:gauge field by stationary particle} does not contain such radiation modes because it is expanded by modes $e^{\mp i\frac{\bm{k}\cdot\bm{p}}{E}t \pm i\bm{k}\cdot\bm{x}}$ which do not follow the dispersion relation of free photons $\omega(\bm{k})=|\bm{k}|$.}, that is, the modes that can be expanded by the Fourier modes $e^{\pm i(-\omega(\bm{k}) t+\bm{k}\cdot \bm{x})}$ with $\omega(\bm{k})=|\bm{k}|$.
In four dimensions, the (sub)leading long-range part of radiations is irrelevant to details of scatterings, and the typical behavior can be studied by considering the bremsstrahlung for the sudden acceleration of a charge \cite{Itzykson:1980rh, Hirai:2020kzx, Hirai:2022yqw}. 
We follow the same steps in three dimensions, and we can find that the radiation modes behave near $\mathscr{I}^+$ as 
\begin{align}
\label{rad-order}
    A^{\mathrm{rad}}_u \sim \order{\frac{1}{\sqrt{r}}},\,
    A^{\mathrm{rad}}_r \sim \order{\frac{1}{r^{\frac{3}{2}}}},\, 
    A^{\mathrm{rad}}_{\theta} \sim \order{\sqrt{r}}.
\end{align}
The detail of the computations is presented in appendix~\ref{app:constraint equations}.
Note that radiation modes \eqref{rad-order} are subdominant compared to the Coulomb potential \eqref{eq:comp_gauge field near I^+}.

Based on this observation, we impose the asymptotic boundary condition on the gauge fields near $\mathscr{I}^+$ as
\begin{align}
\left\{
\begin{alignedat}{7}
    &A_u(u,r,\theta) \\
    &= A_u^{(0,\log)}\log{r} +A_u^{(0)}(\theta)+ A_u^{(-1/2)}(u,\theta)\frac{1}{\sqrt{r}} + \cdots ,\\
    &A_r(u,r,\theta) \\
    &= A_r^{(0,\log)}(\theta)\log{r} +A_r^{(0)}(\theta)+ A_r^{(-1/2)}(\theta)\frac{1}{\sqrt{r}}
    \\
    &\quad+ A_r^{(-1)}(u,\theta)\frac{1}{r} +A_r^{(-3/2)}(u,\theta)\frac{1}{r^{\frac{3}{2}}}+ \cdots, \\
    &A_\theta(u,r,\theta) \\
    &= A_\theta^{(1, \log)}(\theta)r\log{r}+A_\theta^{(1)}(\theta)r + A_\theta^{(1/2)}(u,\theta)\sqrt{r} + \cdots,
    \end{alignedat}
    \right.
    \label{A_exp_I+}
\end{align}
in the Lorenz gauge. 
The first two terms in each component are independent of $u$, represent the Coulomb potential, and do not contain the radiation modes. 
We also allow a $1/\sqrt{r}$ term in $A_r$ although the Coulomb potential \eqref{eq:gauge field by stationary particle} does not contain such a term. 
The coefficient  $A_r^{(-1/2)}$ is $u$-independent and thus it does not contribute to the radiation modes. 
The next order $A_r^{(-1)}$ is allowed to have a $u$-dependence. 
However, it also does not contribute to the radiation modes because the $u$-dependence is completely fixed from other Coulomb potential terms by the Lorenz gauge condition as explained below. 
This is consistent with \eqref{rad-order} where the radiation modes of $A_r$ is $\order{r^{-3/2}}$.
Fields on $\mathscr{I}^-$ have a similar expansion, and we require that they satisfy the antipodal matching condition $A_\mu^{(a)}(u=-\infty, \theta)=A_\mu^{(a)}(v=+\infty, \theta+\pi)$.

Coefficients in the expansion \eqref{A_exp_I+} are related furthermore as
\begin{align}
     &\partial_\theta A_r^{(0,\log)}=A_\theta^{(1, \log)}, \quad 
      \partial_\theta A_r^{(0)}=A_\theta^{(1, \log)}+A_\theta^{(1)},
      \label{no-monopole_cond}
      \\
      &-A_u^{(0,\log)}+A_r^{(0,\log)}+\partial_\theta A_\theta^{(1, \log)}=0,\label{leading_Lorenz}
      \\
      &-A_u^{(0,\log)}+A_r^{(0,\log)}-A_u^{(0)}+A_r^{(0)}+\partial_\theta A_\theta^{(1)}
      \nn
      &=\partial_u  A_r^{(-1)},
      \label{subleading_Lorenz}
\end{align}
where the first two equations \eqref{no-monopole_cond} represent the condition of no long-range magnetic flux $F_{r\theta}=0$ at $\mathscr{I}^+_-$ as imposed in \cite{He:2014cra, Kapec:2015ena}, where $\mathscr{I}^+_\pm$ denotes the sphere at $u=\pm \infty$ in $\mathscr{I}^+$ and similarly $\mathscr{I}^-_\pm$ does the sphere at $v=\pm \infty$.
The last two ones \eqref{leading_Lorenz}, \eqref{subleading_Lorenz} are required from the Lorenz gauge condition.
We suppose below that quantized fields in the covariant gauge satisfy the asymptotic boundary condition \eqref{A_exp_I+}, up to the BRST exact terms.
Note that we have not assumed the confinement to impose \eqref{A_exp_I+}. If we are in the confinement sector, the leading part representing the Coulomb potential modes should vanish.

From \eqref{A_exp_I+}, the field strength has expansions
\begin{align}
\left\{
\begin{alignedat}{3}
    F_{ur}(u,r,\theta) &= F_{ur}^{(-1)}(\theta)\frac{1}{r} + F_{ur}^{(-3/2)}(u,\theta)\frac{1}{r^{3/2}} + \cdots, \\
    F_{r\theta}(u,r,\theta) &=  F_{r\theta}^{(-1/2)}(u,\theta)\frac{1}{\sqrt{r}} + \cdots, \\
    F_{\theta u}(u,r,\theta) &= F_{\theta u}^{(1/2)}(u,\theta)\sqrt{r} + F_{\theta u}^{(0)}(\theta)+\cdots. 
\end{alignedat}
    \right.
    \label{large_Fthetau}
\end{align}
The $u$-independent terms do not contain the radiation modes.
Note that $A_\theta^{(1/2)}$ contributes to $F_{r\theta}^{(-1/2)}, F_{\theta u}^{(1/2)}$, and $A_\theta^{(1/2)}$ gives the leading asymptotic data of radiation.
We will see that the soft parts of our asymptotic charges are indeed determined from $A_\theta^{(1/2)}$.

We consider the asymptotic symmetries in this theory as residual large gauge transformations (modulo the gauge constraints) that preserve the Lorenz gauge and the above boundary conditions.
Let $\alpha(x)$ be a gauge parameter. 
It satisfies $\Box \alpha=0$ to preserve the Lorenz gauge. 
One of such nontrivial large gauge transformations is the one that approaches to $\alpha^{(1/2)}(\theta)\sqrt{r}$ near $\mathscr{I}^+$, where $\alpha^{(1/2)}$ is an arbitrary function on the celestial circle, because it transforms the leading radiation data $A_\theta^{(1/2)}$.
We can find such a solution of $\Box \alpha=0$ in the form 
\begin{align}
\label{def:a-1/2}
    \alpha(x)&= \int d\theta' G^{(1/2)}(x;\theta') \alpha^{(1/2)}(\theta').
\end{align}
See Appendix~\ref{app:LGP} for the explicit form of $G^{(1/2)}(x;\theta')$, where we can also see that the gauge parameters satisfy the antipodal matching condition.
Similarly, we can also find the solution such that it approaches to $\alpha^{(0)}(\theta)$ (see Appendix~\ref{app:LGP}) as
\begin{align}
\label{def:a-0}
    \alpha(x)&= \int d\theta' G^{(0)}(x;\theta') \alpha^{(0)}(\theta'),
\end{align}
although it acts only on the subleading data and does not transform the leading data $A_\theta^{(1/2)}$.\footnote{We may consider larger gauge transformations acting on the Coulomb potential terms, \textit{e.g.} $A_\theta^{(1, \log)}(\theta)$, in \eqref{A_exp_I+}. In this paper, we do not consider them because they are irrelevant to the radiation data.}

\section{Asymptotic charges and confinement in QED}
We construct the asymptotic charges generating the asymptotic symmetries. 
According to Noether's second theorem, the charge associated with the gauge parameter $\alpha(x)$ is given by
\begin{align}
    Q[\alpha] = \int_{\del\Sigma}\alpha*F
\end{align}
where $*$ denotes the Hodge star and $\del\Sigma$ is the one-dimensional (asymptotic) boundary of a Cauchy slice $\Sigma$.
We take the future/past infinities as the Cauchy surfaces since we are interested in the action of asymptotic symmetries on asymptotic states.
The asymptotic charges on the future/past are given as
\begin{align}
    Q^+[\alpha]=\int_{\mathscr{I}^+_-}\alpha*F, \quad 
    Q^-[\alpha]=\int_{\mathscr{I}^-_+}\alpha*F, 
\end{align}
and they are conserved 
\begin{align}
    Q^+[\alpha]=Q^-[\alpha],
\end{align}
supposing that the gauge parameters satisfy the antipodal matching condition.

These charges always vanish for gauge parameters that fall off rapidly at infinity corresponding to small gauge transformations associated with gauge redundancy, while for those related to asymptotic symmetries, they can be non-zero in principle and can generate physical symmetry transformations.

However, we will show that the large gauge transformations \eqref{def:a-1/2} and \eqref{def:a-0} are also trivial under the assumption of confinement.
We use the term \textit{confinement} to mean that asymptotic states do not contain charged particles.
We here only show the explicit expression of the future ones $Q^+$ because the analysis on the past ones $Q^-$ is the same.

First, we consider subleading gauge transformations associated with \eqref{def:a-0}. 
We denote the charges by $Q_0^+$, and they depend on $\alpha^{(0)}$ as
\begin{align}
    Q^+_0[\alpha^{(0)}]&=\lim_{r\to \infty}\int_{\mathscr{I}^+_-}\!\!\!\!\!
    d\theta\, r \alpha^{(0)}(\theta) F_{ru}
    \nn
    &=\int^{2\pi}_0 \!\!  d\theta\, \alpha^{(0)}(\theta) F_{ru}^{(-1)}(\theta).
    \label{Q0_theta-int}
\end{align}
They only depend on the leading Coulombic electric fields $F_{ur}^{(-1)}$, and thus vanish in the sector where no asymptotic charged particles exist. 
%Conversely, if $ Q^+_0[\alpha^{(0)}]$ vanish for arbitrary $\alpha^{(0)}(\theta)$, we can conclude that we are in the sector where $F_{ur}^{(-1)}$ vanishes. 
Since $ Q^+_0[\alpha^{(0)}]$ are independent of radiation data, they do not have the soft parts, unlike QED$_4$ (see also appendix~\ref{app:ASC}). 
We similarly define the past charges $ Q^-_0$, and the conservation law $Q^+_0=Q^-_0$ holds.

Next, we consider leading gauge transformations given by \eqref{def:a-1/2}. 
The charges are given by
\begin{align}
    &\lim_{r\to \infty}\int_{\mathscr{I}^+_-}\!\!\!\!\! d\theta\, r^\frac{3}{2} \alpha^{(1/2)}(\theta) F_{ru}
\end{align}
which have divergent parts schematically written as
\begin{align}
    \lim_{r\to \infty} r^\frac{1}{2}Q^+_0[\alpha^{(1/2)}].
\end{align}
We thus define the charges associated with \eqref{def:a-1/2} by subtracting the divergent parts as
\begin{align}
   & Q^+_{1/2}[\alpha^{(1/2)}]
   :=
   \nn
   &\lim_{r\to \infty}\int^{2\pi}_0 \!\! d\theta\,  \alpha^{(1/2)} \left[r^\frac{3}{2}F_{ru}(u=-\infty,r,\theta)-r^\frac{1}{2}F_{ru}^{(-1)}(\theta)\right]
   \nn
   &=\int_{\mathscr{I}^+_-}\!\!\!\!\! d\theta\,  \alpha^{(1/2)}F_{ru}^{(-3/2)},\label{def:Q12}
\end{align}
and similarly $Q^-_{1/2}$.
This subtraction trick is the same one used in \cite{Campiglia:2016hvg} (see also \cite{Freidel:2019ohg}). $Q^\pm_{1/2}$ are also conserved as long as $Q^\pm_0$ are conserved. 
Furthermore, the subtracted terms are indeed zero in the confinement sector because $Q^\pm_0$ vanish.

Eq.~\eqref{def:Q12} becomes (see appendix~\ref{app:ASC} for details)
\begin{align}
    Q^+_{1/2}[\alpha^{(1/2)}]&=Q^+_{1/2,\rS}[\alpha^{(1/2)}]+Q^+_{1/2,\rH}[\alpha^{(1/2)}],
    \label{eq:Q12_S+H}
\end{align}
where $Q^+_{1/2,\rS/\rH}$ represent soft/hard parts of asymptotic charges defined as
\begin{align}
    Q^+_{1/2,\rS}[\alpha^{(1/2)}]&:=-\int_{\mathscr{I}^+} (\partial_\theta \alpha^{(1/2)})(\partial_u A_\theta^{(1/2)}),
    \label{Q12S}\\
    Q^+_{1/2,\rH}[\alpha^{(1/2)}]&:=\int_{\mathscr{I}^+_+}\!\!\!\!\!d\theta\,  \alpha^{(1/2)}F_{ru}^{(-3/2)}.
    \label{Q12H}
\end{align}
$Q^+_{1/2,\rS}$ has the contribution only from the radiation modes because $ A_\theta^{(1/2)}$ is the leading asymptotic data of radiation. In addition, due to the $u$-integral in \eqref{Q12S}, only soft energy modes of $A_\theta^{(1/2)}$ are relevant to it.

We now show that $Q^+_{1/2,\rS/\rH}$ vanish in the confined sector in QED.
We first consider the soft part $Q^+_{1/2,\rS}$. 
In the asymptotic region, the radiative part of the gauge field has the expansion
\begin{align}
    A^{\mathrm{rad}}_\mu(x)
    &= \int\frac{d^2\bm{k}}{(2\pi)^22|\bm{k}|}[a_\mu(\bm{k})e^{ik\cdot x}+\mathrm{c.c.}]\biggr|_{k^0=|\bm{k}|},
\end{align}
in the covariant gauge.
The canonical quantization leads to the commutation relation
\begin{align}
    [a_\mu(\bm{k}), a^\dagger_\nu(\bm{k'})]=(2\pi)^2 (2|\bm{k}|)\eta_{\mu\nu}\delta^2(\bm{k}-\bm{k'}).
\end{align}
Using the stationary phase approximation as done in e.g. \cite{He:2014cra}, we can evaluate $A^{(1/2)}_\theta$ on $\mathscr{I}^+$ as
\begin{align}
&A^{(1/2)}_\theta(u,\theta)\nn
    &= \int_0^\infty\!\! \frac{dk}{2(2\pi)^{3/2}\sqrt{k}}~\ab[a_\theta(k\hat{x})e^{-iku-i\frac{\pi}{4}}+\mathrm{c.c.}], \label{eq:Fourier expansion of the radiative gauge field}
\end{align}
where 
\begin{align}
  a_\theta(k\hat{x}):=  \partial_\theta \hat{x}^i a_i(k\hat{x}), \qquad \hat{x}(\theta):=\bm{x}/r.
\end{align}
Note that \eqref{eq:Fourier expansion of the radiative gauge field} consists only of the transverse components.
Substituting \eqref{eq:Fourier expansion of the radiative gauge field} into \eqref{Q12S}, we obtain
\begin{align}
\label{Q12s_k-int}
     Q^+_{1/2,\rS}[\alpha^{(1/2)}]&=\frac{1}{2\sqrt{2\pi}}\lim_{k\to 0}\int^{2\pi}_0 \!\!  d\theta
     \,\partial_\theta \alpha^{(1/2)}
     \nn
     &\quad\times\lim_{k\to 0}\sqrt{k}\ab[a_\theta(k\hat{x})e^{-iku+i\frac{\pi}{4}}+\mathrm{c.c.}].
\end{align}
It vanishes because of the factor $\lim_{k\to 0}\sqrt{k}$, unless $a_\theta(k\hat{x})$ has a singular behavior in the soft region $k \sim 0$.
In QED$_4$, as can be seen in the soft photon theorem, $a_\mu(\bm{k})$ indeed gives rise to the soft photon pole $1/|\bm{k}|$ when asymptotic states contain charged particles.
This is the same in three-dimensions, as can be confirmed from computations using Feynman diagrams.
However, in the confinement phase, charged particles cannot appear in asymptotic states or equivalently in external lines.\footnote{Although the classical description of the logarithmic confinement of QED$_3$ may seem to prohibit only charged states, charged particles flying away in different directions also lead to infinite energy even in neutral states.}
Hence, the creation/annihilation operators of soft photons cannot produce any singular behaviors.\footnote{We assume that asymptotic states do not have soft photon dresses that are independent of charged particles. In QED$_4$, we have to put soft photon dresses on asymptotic states in order to obtain IR-safe S-matrix elements. The dresses essentially depend on charged particles because any dressing factor independent of charged particles is irrelevant to scatterings \cite{Hirai:2022yqw}.}
Thus, we conclude that $Q^+_{1/2,\rS}$ vanishes as
\begin{align}
    \bra{\psi}Q^+_{1/2,\rS} = 0
    \label{eq:soft12+0}
\end{align}
for any finite energy final states $\bra{\psi}$. 
A similar discussion holds for the past one $Q^-_{1/2,\rS}$ as $Q^-_{1/2,\rS}\ket{\psi} = 0$.

The hard part $Q^\pm_{1/2,\rH}$ in \eqref{Q12H} also vanishes.
$Q^\pm_{1/2,\rH}$ is determined by the behaviors of electric fields at the time-like infinity. 
In the confinement sector, we do not have any charged fields, and thus there is no Coulomb potential associated with them. 
Thus, the hard part vanishes as $Q_0^\pm$ does. 
One may wonder that the radiative modes might have a non-decaying contribution in the limit $u\to \infty$ because in odd dimensions radiations do not have to propagate in the null direction and can have tails to time-like directions, unlike four-dimensions. 
Nevertheless, the tails vanish in the infinite time limit unless charged particles exist for an infinite time, and radiation modes do not have contributions to $F_{ru}^{(-3/2)}$ at $u=+\infty$ in \eqref{Q12H}. We also confirm this fact explicitly in a typical example in appendix~\ref{app:ASC}.

Therefore, we conclude that in the confinement sector of QED$_3$, the action of the asymptotic charges on any asymptotic states is trivial (or gauged),
\begin{align}
    \bra{\psi}Q^+[\alpha] = 0,\quad   Q^-[\alpha]\ket{\psi'} = 0.
    \label{eq:extended physical condition1}
\end{align}
It means that the large transformations with parameters \eqref{def:a-1/2} and \eqref{def:a-0} are gauge redundancy.

This result stands in contrast to QED$_4$. In the theory, large transformations can be spontaneously broken. 
As a consequence, there exist superselection sectors labeled by the asymptotic charges, and the soft parts of the charges are classically related to a \textit{memory}, i.e. a permanent shift of gauge fields. It is quantum mechanically realized by a coherent cloud of soft photons attached to charged particles \cite{Hirai:2022yqw}.
In QED$_3$, no charged particles can appear on external lines. 
Then, it is reasonable that the hard parts of asymptotic charges vanish in the confinement sector. 
Our analysis further shows that the soft parts also vanish, although photons exist.\footnote{We have not investigated sub-subleading charges that involve higher order terms in the expansion \eqref{large_Fthetau}. It might be possible that they can be non-trivial.}
It means the absence of the memory effect in three-dimensional electromagnetism with finite energy conditions.
It can be understood from the fact that the soft photon theorem is trivial in this theory.
%This conclusion may imply that memory effects do not occur in any confinement theories.

Reversing our argument, we may discuss that \eqref{eq:extended physical condition1} is regarded as a criterion of the confinement; if $Q^\pm_0[\alpha^{(0)}]$, $Q^\pm_{1/2}[\alpha^{(1/2)}]$ act trivial on an asymptotic state for any $\alpha^{(0)}$ and $\alpha^{(1/2)}$, the state does not contain asymptotic charged particles.\footnote{Here the absence of the asymptotic charged particles means that we cannot separate charged particles infinitely. Separation with a finite distance might be allowed.}
Indeed, $Q^\pm_0[\alpha^{(0)}]=0$ for arbitrary functions $\alpha^{(0)}(\theta)$ requires that the asymptotic electric current in each celestial angle should vanish at least at the classical level. In addition, if there is an external charged particle, it produces a $1/|\bm{k}|$ pole for $a_\theta(k\hat{x})$ in \eqref{Q12s_k-int}, and $Q^+_{1/2,\rS}$ then diverges. 
It thus seems that $Q^\pm_{1/2}[\alpha^{(1/2)}]=0$ does not allow asymptotic charged particles. 
We leave for future works the refinement of this discussion to investigate whether the triviality of the action of asymptotic symmetries can generally serve as a criterion for confinement in generic gauge theories. 
In particular, we speculate that the Kugo-Ojima confinement criterion \cite{Kugo:1977zq, Kugo:1979gm} in QCD$_4$ is connected to the triviality of asymptotic symmetries.\footnote{See \cite{Hata:1981nd, Hata:1983cs}, which also discusses ``large gauge symmetry'' in non-abelian gauge theories in a similar spirit.}

%%%%%%%%%%%%%%%%%%
%%%%%%%%%%%%%%%%%%
%%%%%%%%%%%%%%%%%%
\acknowledgments
SS acknowledges support from JSPS KAKENHI Grant Numbers JP21K13927 and JP22H05115.

%%%%%%%%%%%%%%%%%%%%%
%\bibliographystyle{apsrev4-2}
\bibliography{ref}
\bibliographystyle{JHEP.bst}
%%%%%%%%%%%%%%%%%%%%%

%%%%%%%%%%%%%%%%%%
%%%%%%%%%%%%%%%%%%
%%%%%%%%%%%%%%%%%%
\clearpage
\onecolumngrid
\appendix
\begin{center}
\textbf{Supplemental Material}
\end{center}
\section{Radiation and asymptotic boundary conditions}\label{app:constraint equations}

In this section, we confirm a typical radiation behavior 
\begin{align}
\label{rad-order:app}
    A^{\mathrm{rad}}_u \sim \order{\frac{1}{\sqrt{r}}},\,
    A^{\mathrm{rad}}_r \sim \order{\frac{1}{r^{\frac{3}{2}}}},\, 
    A^{\mathrm{rad}}_{\theta} \sim \order{\sqrt{r}}.
\end{align}
and summarize the asymptotic boundary condition.

As in \eqref{eq:comp_gauge field near I^+}, the Coulomb potential part generally has higher order terms.
To be more specific, the potential \eqref{eq:gauge field by stationary particle} for a uniformly moving charge, each component behaves near $\mathscr{I}^+$ as 
\begin{align}
\label{comp_Coulomb}
\left\{
\begin{alignedat}{4}
    A_u &\sim  \frac{e E}{2\pi m}\log\ab(\frac{\varepsilon E (E-\bm{p}\cdot \hat{x})r}{m^2}), \\
    A_r &\sim  \frac{e (E-\bm{p}\cdot \hat{x})}{2\pi m}\log\ab(\frac{\varepsilon E (E-\bm{p}\cdot \hat{x})r}{m^2}), \\
    A_{\theta} &\sim -\frac{er \bm{p}\cdot (\partial_\theta \hat{x})}{2\pi m}\log\ab(\frac{\varepsilon E (E-\bm{p}\cdot \hat{x})r}{m^2})
    =\partial_\theta \left[\frac{er(E-\bm{p}\cdot \hat{x})}{2\pi m}\left(\log\ab(\frac{\varepsilon E (E-\bm{p}\cdot \hat{x})r}{m^2})-1\right)\right].
    \end{alignedat}
    \right.
\end{align}

What contributes to the soft part is a long-range part of the radiation modes, which we will investigate below. 
In four dimensions, the (sub)leading long-range part of radiations is irrelevant to details of scatterings, and the typical behavior can be studied by considering the bremsstrahlung for the sudden acceleration of a charge \cite{Itzykson:1980rh, Hirai:2020kzx, Hirai:2022yqw}. 
We follow the same steps in three dimensions and consider the following current
\begin{align}
    j^\mu(x) = e\sum_{n=I,F}\Theta(\eta_n t)\frac{p_n^\mu}{E_n}\delta^{(2)}\ab(\bm{x}-\frac{\bm{p}_n}{E_n}t),\label{eq:current of kicked particle}
\end{align}
where $\eta_I=-1$ and $\eta_F=1$. 
This current represents the trajectory of a charge kicked at $t=0$, changing the momentum from the initial $p_I^\mu$ to the final $p_F^\mu$. 
We remark that this current cannot be strictly realized with finite energy due to the confinement. 
We will consider a more realistic current in appendix~\ref{app:ASC}. It will turn out that the large $r$-behaviors are the same, although $u$-dependence is different. 
Thus, the following order estimation in large $r$ for the current \eqref{eq:current of kicked particle} provides is sufficient for our purpose of confirmation of \eqref{rad-order:app}.

The Fourier transformation of the current can be computed as
\begin{align}
    \Tilde{j}(k) = -ie\sum_{n=I,F}\frac{\eta_n p_n}{k\cdot p_n - i\eta_n\varepsilon},
\end{align}
using the Fourier transformation of the step function
\begin{align}
    \int dt\ \Theta(\pm t)e^{i\omega t} = \lim_{\varepsilon\to 0+} \frac{\pm i}{\omega\pm i\varepsilon}.
\end{align}
Thus, the gauge field generated by the current can be computed in the Lorenz gauge as
\begin{align}
    A^\mu(x)
    = &e\sum_{n=I,F}\int\frac{d^2\bm{k}}{(2\pi)^2 2|\bm{k}|}\frac{p_n^\mu }{k\cdot p_n}\biggl[\eta_n 
 \Theta(t)\ab(e^{ik\cdot x}+e^{-ik\cdot x})- \Theta(\eta_n t)\ab(e^{-i\frac{\bm{k}\cdot\bm{p}_n}{E_n}t + i\bm{k}\cdot\bm{x}} + e^{i\frac{\bm{k}\cdot\bm{p}_n}{E_n}t - i\bm{k}\cdot\bm{x}})\biggr]\biggr|_{k^0=|\bm{k}|}.
    \label{Amu-kick}
\end{align}
We define the radiation modes as ones with the dispersion relation of free photons, that is, the modes that can be expanded by the Fourier modes $e^{\pm i(-\omega(\bm{k}) t+\bm{k}\cdot \bm{x})}$ with $\omega(\bm{k})=|\bm{k}|$.
The second line in \eqref{Amu-kick} does not have this dispersion relation and represents the Coulomb (or Lienard-Wiechert) potential. 
The first line is the radiation mode for the kicked current, and we represent it by
\begin{align}
A_{\mathrm{rad}}^\mu(x):=e\sum_{n=I,F}\int&\frac{d^2\bm{k}}{(2\pi)^2 (2|\bm{k}|)}\frac{\eta_n  p_n^\mu }{k\cdot p_n}
\Theta(t)\ab(e^{ik\cdot x}+e^{-ik\cdot x})\biggr|_{k^0=|\bm{k}|}.
\end{align}

The behavior near $\mathscr{I}^+$ is evaluated by the stationary phase approximation as\footnote{We take the contributions from the stationary point $\theta=0$ and drop the ones from another stationary point $\theta=\pi$ as, e.g., in \cite{Strominger:2017zoo}.} 
\begin{align}
A_{\mathrm{rad}}^\mu(x)&= \frac{e}{2\sqrt{r}}\sum_{n=I,F}\frac{\eta_n  p_n^\mu }{-E_n +\bm{p}_n \cdot \hat{x}}
\int^\infty_0\!\!\!\frac{d k}{(2\pi k)^\frac{3}{2}}
\ab(e^{-ik u-\frac{i\pi}{4}}+\mathrm{c.c.})+\mathcal{O}(r^{-\frac{3}{2}}).
\label{eq:gauge field by kicked instantaneously after saddle point approx}
\end{align}
Unlike the four-dimensional case, the remaining $k$-integral has an IR divergence near $k \sim 0$ reflecting the fact that the current \eqref{eq:current of kicked particle}, where charged particles exist for infinite time, cannot be strictly realized due to the logarithmic confinement as noted above. 
Nevertheless, it does not affect the $r$-dependence, and we conclude the radiation mode behaves near $\mathscr{I}^+$ as \eqref{rad-order:app}.
Note that $1/\sqrt{r}$ order in $A^{\mathrm{rad}}_r$ vanishes\footnote{It vanishes even for more general kick process considered in \cite{Hirai:2020kzx}, when we have the charge conservation.} due to 
\begin{align}
    \sum_{n=I,F}\frac{\eta_n  p_{nr} }{-E_n +\bm{p}_n \cdot \hat{x}}=\sum_{n=I,F}\eta_n=0,
\end{align}
and it is also required from $F_{ur}=\mathcal{O}(1/r)$ which is the finiteness condition of electric flux discussed below.
The radiation modes \eqref{rad-order:app} are subdominant compared to the Coulomb potential \eqref{comp_Coulomb}.

We can also see that the behavior \eqref{large_Fthetau}, which is derived from the asymptotic boundary condition \eqref{A_exp_I+}, is consistent with physical conditions as follows.
Since the energy flux passing through the null infinity whose surface area grows like $r$ should be finite, we must have
\begin{align}
    T_{uu}\sim\order{\frac{1}{r}}. \label{eq:finiteness of energy flux}
\end{align}
Similarly, from the finiteness of physical quantities such as the current, the total electric charge, the magnetic charge, we find
\begin{align}
    j_u\sim\order{\frac{1}{r}},\quad F_{ur} \sim \order{\frac{1}{r}}, \quad F_{r\theta} \sim \order{1}. \label{eq:finiteness of charge}
\end{align}
Combining them with the Bianchi identity, we obtain the general large-$r$ expansion of the field strength near $\mathscr{I}^+$ as 
\begin{align}
    F_{ur}(u,r,\theta) &= F_{ur}^{(-1)}(u,\theta)\frac{1}{r} + \cdots, \\
    F_{r\theta}(u,r,\theta) &=  F_{r\theta}^{(0)}(u,\theta) + \cdots, \\
    F_{\theta u}(u,r,\theta) &= F_{\theta u}^{(1/2)}(u,\theta)\sqrt{r}+\cdots .
\end{align}
If we further require that there is no long-range magnetic flux as in the main text, we should have $ F_{r\theta}^{(0)}(u,\theta)=0$, and then the large-$r$ behavior of the field strength is consistent with \eqref{large_Fthetau}.

\section{Large gauge parameters}
\label{app:LGP}
In this section, we provide the explicit expressions of the large gauge parameters used in the main text.

The residual gauge parameters satisfying $\Box \alpha(x)=0$ such that it asymptotically approaches to $\alpha^{(1/2)}(\theta)\sqrt{r}$ in the limit that $x$ reaches $\mathscr{I}^+$, where $\alpha^{(1/2)}$ is an arbitrary function on the celestial circle, is given as
\begin{align}
    \alpha(x)&= \int d\theta' G^{(1/2)}(x;\theta') \alpha^{(1/2)}(\theta'),
    \label{a-1/2}
    \\
    G^{(1/2)}(x;\theta')&:=\lim_{\epsilon\to 0+}\left[\frac{(-x\cdot x+2i(u+r)\epsilon)}{8\sqrt{2}[-q(\theta')\cdot x+i\epsilon]^\frac{3}{2}}+\mathrm{c.c.}\right],
    \label{G-1/2}
\end{align}
where $q^\mu(\theta')=(1, \cos\theta',\sin\theta')$. 
Of course, the solutions are not unique in the sense that we have the freedom to add subleading terms that do not change the leading behavior $\alpha^{(1/2)}(\theta)\sqrt{r}$.
Thus, \eqref{G-1/2} is just one of the solutions. 
$\epsilon$ is introduced to avoid the branch cuts by shifting $x^\mu=(t, r \hat{x})\to (t\pm i \epsilon, r \hat{x})$ in the Cartesian coordinate.
It is equivalent to shifting $u$ as $u\pm i\epsilon$ in the retarded coordinates.
In the limit approaching the past null infinity $\mathscr{I}^-$, $\alpha(x)$ asymptotically approaches to $\alpha(x) \to \alpha^{(1/2)}(\theta+\pi)\sqrt{r}$.
Thus, the gauge parameters satisfy the antipodal matching condition
\begin{align}
    \lim_{u\to-\infty}\lim_{r\to\infty}\alpha(t,r\hat{x}) = \lim_{v\to\infty}\lim_{r\to\infty}\alpha(t,-r\hat{x}).
\end{align}

Similarly, the residual gauge parameters asymptotically approaching to $\alpha^{(0)}(\theta)$ in the limit approaching $\mathscr{I}^+$, where $\alpha^{(0)}$ is an arbitrary function on the celestial circle, is given as
\begin{align}
    \alpha(x)&= \int d\theta' G^{(0)}(x;\theta') \alpha^{(0)}(\theta'),\\
    G^{(0)}(x;\theta')&:=\lim_{\epsilon\to 0+}\left[\frac{\sqrt{-x\cdot x+2i(u+r)\epsilon}}{4\pi(-q(\theta')\cdot x+i\epsilon)}+\mathrm{c.c.}\right].
\end{align}
It also satisfies the antipodal matching condition.

\section{Notes on asymptotic charges}
\label{app:ASC}
We here summarize some results of asymptotic charges that we use in the main text.

In order to decompose the asymptotic charges into the hard and soft parts, we use a constrained equation:
\begin{align}
    -j_u&=j^u+j^r=\nabla_\mu F^{u\mu}+\nabla_\mu F^{r\mu}
    \nn
    &=-\partial_u F_{ru}+\frac{1}{r}\partial_r\left(r F_{ru}\right)-\frac{1}{r^2}\partial_\theta F_{u\theta}.
\end{align}
It leads to 
\begin{align}
    \partial_u F_{ru}=j_u+\frac{1}{r}\partial_r\left(r F_{ru}\right)-\frac{1}{r^2}\partial_\theta F_{u\theta}.
\end{align}
Substituting the expansion \eqref{large_Fthetau}, we obtain, e.g., 
\begin{align}
\partial_u F_{ru}^{(-1)}=j_u^{(-1)}, \quad
\partial_u F_{ru}^{(-3/2)}=j_u^{(-3/2)}-\partial_\theta F_{u\theta}^{(1/2)}
\end{align}
on $\mathscr{I}^+$, where $j_u^{(a)}$ represent coefficients of $\mathcal{O}(r^{a})$ terms in the large $r$ expansion of $j_u$. 
If the theory does not contain massless charged particles, $j_u^{(-1)}$ and $j_u^{(-3/2)}$ vanish because $j_u$ decays faster than $r^{-1}$. Intuitively, we can understand it from the fact that massive particles reach the time-like infinity rather than the null infinity.
This fact can also be seen explicitly by expanding a massive matter field, say a Dirac field, in the Fourier modes and checking that it has no saddle points for large-$r$ with $u$ fixed.
We therefore have
\begin{align}
\label{constraint-F}
\partial_u F_{ru}^{(-1)}=0, \quad 
    \partial_u F_{ru}^{(-3/2)}=-\partial_\theta F_{u\theta}^{(1/2)}.
\end{align}

Let us rewrite the asymptotic charges using the above equations.
We first consider $Q_0$ given in \eqref{Q0_theta-int}. 
It is given by
\begin{align}
    Q^+_0[\alpha^{(0)}]&=\int^{2\pi}_0 \!\!  d\theta\, \alpha^{(0)}(\theta) F_{ru}^{(-1)}(u=-\infty,\theta)
    \nn
    &=-\int^\infty_{-\infty}\!\!\!d u\int^{2\pi}_0 \!\!  d\theta\, \alpha^{(0)}(\theta) \partial_u F_{ru}^{(-1)}(u,\theta)+\int^{2\pi}_0 \!\!  d\theta\, \alpha^{(0)}(\theta) F_{ru}^{(-1)}(u=+\infty,\theta).
    \label{eq:Q0-u-int}
\end{align}
The first term vanishes by \eqref{constraint-F}. 
We thus have
\begin{align}
    Q^+_0[\alpha^{(0)}]
    &=\int^{2\pi}_0 \!\!  d\theta\, \alpha^{(0)}(\theta) F_{ru}^{(-1)}(u=+\infty,\theta).
\end{align}
In other words, since $F_{ru}^{(-1)}(\theta)$ is independent of $u$, we can just write
\begin{align}
        Q^+_0[\alpha^{(0)}]
    &=\int^{2\pi}_0 \!\!  d\theta\, \alpha^{(0)}(\theta) F_{ru}^{(-1)}(\theta).
\end{align}
We can also say that $Q_0$ does not have the soft part because the $u$-integral part (the first term in \eqref{eq:Q0-u-int}) vanishes.

We can also see that the hard part of $Q_0$ vanishes in the confinement sector.
$F_{ru}^{(-1)}(\theta)$ contains only the Coulomb modes as explained in the main text. We here evaluate it for a case such that there are outgoing charged particles (labeled by $n$) with mass $m_n$, momentum $p_n^\mu$ and charge $e_n$ in the asymptotic future.
The Coulomb potential is given as
\begin{align}
    &A^\mu(x) 
     = - \sum_n \frac{e_n p_n^\mu}{2\pi m_n} \log\ab(\frac{\varepsilon E_n\sqrt{(p_n \cdot x)^2 + m_n^2 (x \cdot x)}}{m_n^2}).
\end{align}
It leads to
\begin{align}
    F_{ur}^{(-1)}(\theta)=-\sum_n\frac{e_n m_n}{2\pi (E_n-\bm{p}_n\cdot \hat{x}(\theta))}.
\end{align}
In particular in the confinement sector, there are no these asymptotic charged particles, and thus we have $ F_{ur}^{(-1)}(\theta)=0$. It means $Q_0^+=0$.
Conversely, if we have $Q_0^+[\alpha^{(0)}]=0$ for arbitrary $\alpha^{(0)}(\theta)$, we can conclude that there are no asymptotic charged particles (assuming all asymptotic charged particles have different velocities).

We next consider $Q_{1/2}$ in \eqref{def:Q12}. 
It is rewritten as 
\begin{align}
    Q^+_{1/2}[\alpha^{(1/2)}]&=\int^{2\pi}_0 \!\!  d\theta\, \alpha^{(1/2)}(\theta) F_{ru}^{(-3/2)}(u=-\infty,\theta)
    \nn
    &=-\int^\infty_{-\infty}\!\!\!d u\int^{2\pi}_0 \!\!  d\theta\, \alpha^{(1/2)}(\theta) \partial_u F_{ru}^{(-3/2)}(u,\theta)+\int^{2\pi}_0 \!\!  d\theta\, \alpha^{(1/2)}(\theta) F_{ru}^{(-3/2)}(u=+\infty,\theta)
    \nn
    &=\int^\infty_{-\infty}\!\!\!d u\int^{2\pi}_0 \!\!  d\theta\, \alpha^{(1/2)}(\theta) \partial_\theta F_{u\theta}^{(1/2)}(u,\theta)+\int^{2\pi}_0 \!\!  d\theta\, \alpha^{(1/2)}(\theta) F_{ru}^{(-3/2)}(u=+\infty,\theta),
    \label{eq:Q12-u-int}
\end{align}
where we have used \eqref{constraint-F}.
The asymptotic boundary condition further leads to 
$F_{u\theta}^{(1/2)}=\partial_u A_\theta^{(1/2)}$.
We thus obtain \eqref{eq:Q12_S+H} with \eqref{Q12S} and \eqref{Q12H}.

We now confirm that the hard part \eqref{Q12H} of the asymptotic charge $Q^+_{1/2}$ vanishes in the confinement sector.
The absence of contributions from the Coulomb modes is similar to the above discussion of $Q^+_{0}$. 
Below, we see that the radiation modes also do not contribute to \eqref{Q12H}.
The example with current \eqref{eq:current of kicked particle} is not appropriate to this aim because the charged particles exist in the asymptotic region. 
We thus consider the following example where charged particles exist only for a finite time, and explicitly demonstrate the vanishing of contributions from the radiation modes. 
The current, which mimics a particle-antiparticle creation and annihilation process, is depicted as
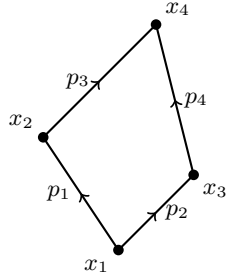
\begin{center}
    \begin{tikzpicture}[every node/.style={font=\small},
        arrowstyle/.style={
            postaction={decorate,
                decoration={
                    markings,
                    mark=at position 0.5 with {\arrow{>}}
                }
            }
        }
    ]
        % 各頂点の座標を定義
        \coordinate (A) at (2,0);
        \coordinate (B) at (3,1);
        \coordinate (C) at (2.5,3);
        \coordinate (D) at (1,1.5);

        \fill (A) circle (2pt);
        \fill (B) circle (2pt);
        \fill (C) circle (2pt);
        \fill (D) circle (2pt);
        
        % 各辺を描き、辺の中央に矢印とラベルを配置
        \draw[thick, arrowstyle] (A) -- (B) node[midway, right] {$p_2$};
        \draw[thick, arrowstyle] (B) -- (C) node[midway, right] {$p_4$};
        \draw[thick, arrowstyle] (D) -- (C) node[midway, left] {$p_3$};
        \draw[thick, arrowstyle] (A) -- (D) node[midway, left] {$p_1$};
        
        % 各頂点にラベルを配置
        \node[below left] at (A) {$x_1$};
        \node[below right] at (B) {$x_3$};
        \node[above right] at (C) {$x_4$};
        \node[above left] at (D) {$x_2$};
    \end{tikzpicture}
\end{center}
where a particle with charge $+e$ goes along the left path ($x_1\to x_2\to x_4$) and an antiparticle with charge $-e$ along the right path  ($x_1\to x_3\to x_4$). 
First of all, let us evaluate the radiation from one edge of the above paths described as
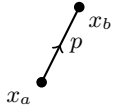
\begin{center}
    \begin{tikzpicture}[every node/.style={font=\small},
        arrowstyle/.style={
            postaction={decorate,
                decoration={
                    markings,
                    mark=at position 0.5 with {\arrow{>}}
                }
            }
        }
    ]
        \coordinate (A) at (2,0);
        \coordinate (B) at (2.5,1);
        \draw[thick, arrowstyle] (A) -- (B) node[midway, right] {$p$};
        \fill (A) circle (2pt);
        \fill (B) circle (2pt);
        \node[below left] at (A) {$x_a$};
        \node[below right] at (B) {$x_b$};
    \end{tikzpicture}
\end{center}
or equivalently
\begin{align}
    j^\mu(x) = q\Theta(t-t_a)\Theta(t_b-t)\frac{p^\mu}{E}\delta\ab(\bm{x}-\frac{\bm{p}}{E}(t-t_a)-\bm{x}_a),
\end{align}
where $q$ takes the value $\pm e$ corresponding to the choice of edges.
The electromagnetic potential is given by
\begin{align}
    A^\mu(x)
    &= \int d^3x'~G(x-x')j^\mu(x').
\end{align}
Since in the momentum space the current can be written as
\begin{align}
    \Tilde{j}_\mu(k)
    &= q\frac{ip_\mu}{p\cdot k}\ab[e^{-ik\cdot x_b}-e^{-ik\cdot x_a}],
\end{align}
the radiation modes are given by
\begin{align}
    A^\mu_\text{rad}(x)
    = q \int\frac{d^2\bm{k}}{(2\pi)^22|\bm{k}|}\ab\{\ab[\frac{p^\mu}{p\cdot k}e^{ik\cdot x}\ab(e^{-ik\cdot x_b}-e^{-ik\cdot x_a})]\biggr|_{k^0=|\bm{k}|} + \text{c.c.}\}
\end{align}
where we do not consider the pole corresponding to $p\cdot k=0$ since we are interested only in the radiation modes here.
We have also assumed that $t$ is much larger than each $t_a$, which is always satisfied if we only consider the radiation near $\mathscr{I}^+_+$.

In order to evaluate the behavior near $\mathscr{I}^+$, we use the saddle point approximation as
\begin{align}
    \int_{y_i}^{y_f} dy~ f(y)e^{-rg(y)} = e^{-rg_s}\biggl(f_s\sqrt{\frac{2\pi}{g_s''}}r^{-1/2} +\sqrt{\frac{\pi}{2}}\ab(\frac{f_s''}{\sqrt{g_s''^3}} 
    -\frac{f_s'g_s^{(3)}}{\sqrt{g_s''^5}} 
    - \frac{f_sg_s^{(4)}}{4\sqrt{g_s''^5}} 
    +\frac{5 f_s(g_s^{(3)})^2}{12 \sqrt{g_s''^7}} )r^{-3/2}
    + \order{r^{-5/2}}\biggr).
\end{align}
Notations used above are defined as
\begin{align}
    y_s : \text{saddle\ of}\ g(y),\quad
    f_s = f(y_s), \quad
    f_s^{(n)} = \frac{d^n}{dy^n}f(y)\biggr|_{y=y_s}.
\end{align}
We have also assumed $y_i<y_s<y_f$.
In our case, $f,g$ are replaced by
\begin{align}
    f(y) \to e^{-i|\bm{k}|u}\frac{p_i^\mu}{p_i\cdot k}e^{-ik\cdot x_b}, \quad
    g(y) \to  i|\bm{k}|(1-\cos{\theta})
\end{align}
respectively, where $\theta$ (which is the angle between $\bm{k}$ and $\bm{x}$) corresponds to the integration variable $y$ in the above saddle point approximation formula. 
Here $b$ denotes $1,\cdots,4$,
and each variable is given by
\begin{gather}
    y_s = \theta_s =0,\quad g_s=g_s'=0,\quad g_s'' = i|\bm{k}|,\quad g_s^{(3)}=0,\quad g_s^{(4)}=-i|\bm{k}|.
\end{gather}
Thus, to the order of $r^{-3/2}$, we have
\begin{align}
\label{Aij_r-exp}
    &\int\frac{d^2\bm{k}}{(2\pi)^2 2|\bm{k}|}e^{-i|\bm{k}|u}\frac{p_i^\mu}{p_i\cdot k}e^{-ik\cdot x_b}e^{-r\cdot i|\bm{k}|(1-\cos{\theta})} \notag\\
    &= \frac{p_i^\mu}{-E_i + \bm{p}_i\cdot \hat{x}}\ab[\int_0^\infty \frac{d|\bm{k}|}{2(2\pi)^{3/2}}\frac{e^{-i|\bm{k}|u-i|\bm{k}|(-t_b+\hat{x}\cdot\bm{x}_b))}}{(i|\bm{k}|)^{1/2}|\bm{k}|}] \times r^{-1/2} \notag \\
    &\quad + \biggl[\frac{p_i^\mu}{(-E_i+\bm{p}_i\cdot \hat{x})^3}\int_0^\infty \frac{d|\bm{k}|}{4(2\pi)^{3/2}}\frac{e^{-i|\bm{k}|u-i|\bm{k}|(-t_b+\hat{x}\cdot\bm{x}_b))}}{(i|\bm{k}|)^{3/2}|\bm{k}|}( \alpha(p_i, \hat{x})+|\bm{k}|\beta(p_i,\bm{x}_b, \hat{x})+|\bm{k}|^2\gamma(\bm{x}_b, \hat{x})) \biggr] \times r^{-3/2} \notag\\
    &\quad + \order{r^{-5/2}},
\end{align}
where 
%$\alpha$ is a function of $p_i,\hat{x}$, $\beta$ is a function of $p_i,\bm{x}_b,\hat{x}$ and $\gamma$ is a function of $\bm{x}_b,\hat{x}$. 
$\alpha, \beta, \gamma$ are some functions with the stated arguments, and 
the detailed expressions of them are unnecessary in the following analysis. 
We represent \eqref{Aij_r-exp} by $A_{i,b}^\mu(x)$ below.

We have obtained the radiation from the original paths as
\begin{align}
    A^\mu_\text{rad}(x) = e\sum_{i,b}\eta_{ib}(A_{i,b}^\mu(x)+\text{c.c.})
\end{align}
where $\eta_{ib}$ is given by
\begin{align}
    \eta_{11}=\eta_{32}=\eta_{23}=\eta_{44}=-1, \quad
    \eta_{12}=\eta_{34}=\eta_{21}=\eta_{43}=1.
\end{align}

What is of interest now is $F_{ru}^{(-3/2)}$ because it constitutes the hard part \eqref{Q12H} of the asymptotic charge as $Q_{1/2,H}^+[\alpha^{(1/2)}]=\int_{\mathscr{I}^+_+}d\theta~\alpha^{(1/2)}F_{ru}^{-3/2}$.
This component of the field strength can be written in terms of the gauge field as
\begin{align}
\label{eq:field strength of -3/2 order in app}
    F_{ru}^{(-3/2)} = \del_uA^{u,(-3/2)} + \frac{1}{2}A^{u,(-1/2)} + \frac{1}{2}A^{r,(-1/2)}.
\end{align}
First of all, note that the second term vanishes since the factor in front of the leading term is independent of the paths, as
\begin{align}
    \frac{p^u_i}{-E_i+\bm{p}_i\cdot\hat{x}} = \frac{E_i-\hat{x}\cdot \bm{p}_i}{-E_i+\bm{p}_i\cdot\hat{x}} = -1,
\end{align}
and this independence leads to the cancellation of the contributions of all paths due to $\eta_{ia}$.
The first and third terms of \eqref{eq:field strength of -3/2 order in app} are given by
\begin{align}
\label{eq:u-deriv of A^u in app}
    &\del_uA^{u,(-3/2)}\nn
    &=  \sum_{i,b}\eta_{ib}\frac{ie}{(-E_i+\bm{p}_i\cdot \hat{x})^2}\int_0^\infty \frac{d|\bm{k}|}{4(2\pi)^{3/2}}\frac{e^{-i|\bm{k}|u-i|\bm{k}|(-t_b+\hat{x}\cdot\bm{x}_b))}}{(i|\bm{k}|)^{3/2}} ( \alpha(p_i, \hat{x})+|\bm{k}|\beta(p_i,\bm{x}_b, \hat{x})+|\bm{k}|^2\gamma(\bm{x}_b), \hat{x})+ \text{c.c.}
\end{align}
and
\begin{align}
\label{eq:A^r in app}
    \frac{1}{2}A^{r,(-1,2)} = \sum_{i,b}e\eta_{ib}\frac{\bm{p}_i\cdot\hat{x}}{-E_i + \bm{p}_i\cdot \hat{x}}\int_0^\infty \frac{d|\bm{k}|}{4(2\pi)^{3/2}}\frac{1}{(i|\bm{k}|)^{1/2}|\bm{k}|}e^{-i|\bm{k}|u-i|\bm{k}|(-t_b+\hat{x}\cdot\bm{x}_b))}+\text{c.c.},
\end{align}
respectively. 
In fact, these terms vanish in the limit of $u\to\infty$ since the region around $|\bm{k}|\sim0$ dominates in this limit and all integrands, behaving as $\order{|\bm{k}|^{-1/2}}$ near $|\bm{k}|=0$,  do not lead to any divergences.
Indeed, each term in the integrand behaves as $|\bm{k}|^{-3/2}$, $|\bm{k}|^{-1/2}$ or higher-order terms in $|\bm{k}|$ as $|\bm{k}|\sim 0$.
All terms behaving as $|\bm{k}|^{-3/2}$, such as the term including $\alpha$ in the first integral of \eqref{eq:u-deriv of A^u in app} and the integrand of \eqref{eq:A^r in app}, depend on $b$ only through the factor $e^{-i|\bm{k}|(-t_b+\hat{x}\cdot\bm{x}_b))}$ and, after summation over $b$ with fixed $i$, this dependence yields $\order{|\bm{k}|}$ as
\begin{align}
    \sum_{b}\eta_{ib}e^{-i|\bm{k}|(-t_b+\hat{x}\cdot\bm{x}_b))} \sim e^{-i|\bm{k}|(-t_b+\hat{x}\cdot\bm{x}_b))}-e^{-i|\bm{k}|(-t_{b'}+\hat{x}\cdot\bm{x}_{b'}))} \sim |\bm{k}|\quad (|\bm{k}|\to0).
\end{align}
%$e^{-i|\bm{k}|(-t_b+\hat{x}\cdot\bm{x}_b))}-e^{-i|\bm{k}|(-t_{b'}+\hat{x}\cdot\bm{x}_{b'}))}\sim |\bm{k}|$ as $|\bm{k}|\to 0$.
Consequently, the integrands behave as $|\bm{k}|^{-1/2}$ near $|\bm{k}|=0$.

After all, since we have confirmed that $F_{ur}^{-3/2}$ vanishes in the limit of $u\to\infty$, we can conclude  that $Q_{1/2,H}^+[\alpha^{(1/2)}]=\int_{\mathscr{I}^+_+}d\theta~\alpha^{(1/2)}F_{ru}^{(-3/2)}$ also disappears.
\end{document}